\documentclass[aps,preprint,a4paper,showpacs,showkeys,superscriptaddress,nofootinbib]{revtex4-1}
\usepackage{latexsym}
\usepackage{amsmath,amssymb}
\usepackage{graphicx}
\usepackage{subfigure}
\usepackage{xcolor}
\usepackage{cancel}
\usepackage{multirow,tabularx,boldline,array}

\newcolumntype{C}[1]{>{\centering\arraybackslash}p{#1}}
\usepackage{hyperref}     % color hypertext for pdflatex
\hypersetup{colorlinks,%
  citecolor=blue,%
  linkcolor=cyan,%
  pdftex}

\usepackage[titletoc]{appendix}
\usepackage{enumerate}
\begin{document}

%\preprint{arXiv:yymm.nnnn [gr-qc]}

\title{Analytic approach to the formation of a three-dimensional black string from a dust cloud}

\author{Hwajin Eom}%
\email[]{um16@sogang.ac.kr}%
\affiliation{Research Institute for Basic Science, Sogang University, Seoul, 04107, Republic of Korea}%
%\affiliation{Department of Physics, Sogang University, Seoul, 04107, Republic of Korea}%
%\affiliation{Center for Quantum Spacetime, Sogang University, Seoul 04107, Republic of Korea}%

\author{Wontae Kim}%
\email[]{wtkim@sogang.ac.kr}%
\affiliation{Department of Physics, Sogang University, Seoul, 04107, Republic of Korea}%
\affiliation{Center for Quantum Spacetime, Sogang University, Seoul 04107, Republic of Korea}%

\date{\today}

\begin{abstract}
In three-dimensional low-energy string theory,
we study the formation of a black string from a dust cloud.
We analytically obtain two distinct classes of exact solutions with arbitrary functions
responsible for mass distributions of
the dust cloud. The first and second kinds of solutions may describe collapsing dusts but
the first kind is only for inhomogeneous dust distribution while the second kind has a homogeneous limit.
The finite collapse time and the Israel junction conditions tell us that
the first kind solution describes
a desired collapsing phenomenon, whereas
the scale factor in the inner spacetime for the second kind
turns out to be trivial.
In the first kind solution, specific collapsing models can be realized by choosing an
appropriate inhomogeneous
dust distribution consistent with the Israel junction conditions.
Consequently, the inhomogeneous dust cloud eventually collapses to the
black string although the homogeneous dust cloud
does not guarantee the formation of the black string in our setting.
The space-like curvature singularities occur at the finite collapse time and
they can be cloaked by the horizon of the black string.
\end{abstract}

%

% \pacs{04.70.Dy, 04.62.+v, 04.60.Kz }

\keywords{Black strings, Gravitational collapse, Junction conditions, Inhomogeneous dust, Cylindrical symmetry}

\maketitle
%%%%%%%%%%%%%%
%% RevTeX Style End %%
%%%%%%%%%%%%%%

%\newcommand{\lp}{\ell_P}

\raggedbottom

\section{introduction}
\label{sec:introduction}

Gravitationally collapsing objects such as massive stars are one of the fundamental subjects in general relativity.
The first theoretical approach to collapsing stars was taken by Oppenheimer and Snyder in 1939,
solving gravitational field equations for a spherically symmetric and spatially homogeneous dust cloud \cite{Oppenheimer:1939ue}.
From the exact solution of the equations, they showed that a Schwarzschild black hole is formed at the end of the collapsing.
Thanks to its simplest symmetry, the collapse of spherically symmetric objects
in four-dimensional spacetime has
been extensively studied in Refs.~\cite{Bondi:1964zza,Penrose:1964wq,Joshi:1994br,Kofinas:2004pt,Cai:2005ew,Apostolopoulos:2006eg,Bronnikov:2008by,Rahaman:2011cw,Frolov:2015bia,Eby:2016cnq,Helfer:2016ljl,Dalianis:2018frf}.

To generalize the geometry of collapsing objects,
cylindrically symmetric spacetime has also been investigated.
A pioneering work on the gravitational collapse in this spacetime is
about
the evolution of an infinitely long cylindrical shell \cite{thorne1972nonspherical}.
Since then,
various types of matter for cylindrical collapsing objects have been considered,
for example,
a thin shell of dust \cite{Apostolatos:1992qqj,Echeverria:1993wf,Lemos:1998iy},
a homogeneous dust cloud \cite{Senovilla:1997zz,Lemos:1997bd,Tod:2004hb},
some fluids \cite{DiPrisco:2009zc,Greenwood:2009gp} and a massless scalar field \cite{Wang:2003vf}.
In particular, in Refs.~\cite{Lemos:1997bd,Senovilla:1997zz,Tod:2004hb,DiPrisco:2009zc,Greenwood:2009gp}, the formation of a four-dimensional black string
has been studied, in which a singularity is surrounded by the horizon of the black
string.
However, unlike these four-dimensional cases, cylindrical collapse to form black strings
has rarely been discussed in lower dimensions.

In three dimensions, the non-rotating Bañados-Teitelboim-Zanelli(BTZ) black hole \cite{Banados:1992wn} can be formed through
homogeneous \cite{Ross:1992ba} and inhomogeneous dust clouds \cite{Gutti:2005pn}.
Also, fluids with nonzero pressure
have been considered for the formation of BTZ black holes, for example,
perfect fluids \cite{Cruz:1994ar,Sharma:2011iq,GarciaDiaz:2014gqn},
a rotating star \cite{Lubo:1998ue} and
a polytropic star \cite{Sa:1999yf}.
On the other hand, Horowitz and Welch
found a black string solution
dual to BTZ black hole solution \cite{Horowitz:1993jc}\footnote[1]{Several black string solutions in higher dimensions are found in general relativity \cite{Mann:2006yi,Cisterna:2017qrb}.}.
Motivated by the formation of the non-rotating BTZ black hole,
one can also consider the formation of the uncharged black string from the dual point of view.
It was claimed that the black string
could also be formed from a dust cloud in three dimensions
from the numerical argument on the dust edge \cite{Hyun:2006xt}.

In this paper,
we will revisit the formation of a static uncharged black string from a
collapsing dust cloud in three-dimensional spacetime
and try to obtain exactly soluble analytic solutions describing the collapsing dust cloud.
If one were to match the inner spacetime of the dust cloud with the outer spacetime of the black string,
the Israel junction conditions would require that
the dust distribution should be spatially inhomogeneous; otherwise,
the initial radius of the dust cloud should be inside the event horizon of the black string.
Based on this fact,
we will analytically obtain collapsing solutions describing the spatially inhomogeneous dust cloud and
discuss physical consequences of them.

The organization of this paper is as follows.
In Sec.~\ref{sec:junction}, we will find the Israel junction conditions
and explain the reason why the dust could should be
inhomogeneous.
Next, in Sec.~\ref{sec:inner},
we will obtain two types of exact solutions describing the dust cloud
and study their properties. The solutions have unspecified local functions
responsible for the dust distribution.
Then, in Sec.~\ref{sec:matching},
we will investigate the inner structure of the dust cloud
by choosing the unspecified local functions appropriate to the junctions conditions.
In Sec.~\ref{sec:simple}, we will present the exact collapsing solution of the inhomogeneous dust, which is smoothly connected with
the geometry of the black string. In finite collapse time,
the dust eventually forms the black string, and the curvature
singularity appears at the end of the collapsing.
Finally, conclusion and discussion will be given in Sec.~\ref{sec:conclusion}.

\section{Israel junction conditions for a dust cloud}
\label{sec:junction}
Let us start with the low-energy effective action in string theory
    \begin{equation}
    \label{eq:action}
    S =\frac{1}{2 \kappa^2} \int d^3 x \sqrt{-g} e^{-2\phi} \left( g^{\mu\nu} R_{\mu\nu}+4\left(\nabla\phi\right)^2 +\frac{4}{\ell^2} \right)+S_M,
    \end{equation}
where $\kappa^2$ is a parameter of length scale, $\phi$ is a dilaton field, $\ell$ is the radius of anti-de Sitter space
and $S_M$ is a matter action. Note that we set $\kappa^2=1$ for simplicity.
The equations of motions with respect to $g_{\mu\nu}$ and $\phi$ are obtained as
    \begin{gather}
    \label{eq:eom_total1}
    e^{-2\phi} \left(R_{\mu\nu}+2\nabla_\mu \nabla_\nu \phi\right)=T^M_{\mu\nu},\\
    \label{eq:eom_total2}
    g^{\mu\nu} R_{\mu\nu}-4\left(\nabla\phi\right)^2 + 4\Box\phi+\frac{4}{\ell^2} =0,
    \end{gather}
in which $T^M_{\mu\nu} =\left(-2/\sqrt{-g} \right) \left(\delta S_M /\delta g^{\mu\nu}\right)$.

In order to investigate a collapsing object in the context of Israel's formulation \cite{Israel:1966rt,poisson_2004},
a hypersurface describing the edge of a dust cloud is defined at $L=0$,
and thus $L>0$ and $L<0$ correspond to the outer and inner spacetimes of the dust cloud, respectively.
Note that $L$ is the radius of the outer spacetime from the edge, $i.e.$, $L= R-\mathcal{R}(t)$,
where $R$ is an outer radial coordinate and $\mathcal{R}(t)$ is the location of the edge with a comoving time $t$.
Using the unit step function defined as $\Theta(L)=1$ for $L>0$ and $\Theta(L)=0$ for $L<0$,
one can write the metric tensor and the dilaton field such as
$g_{\mu\nu} = \Theta(L) g^{\textrm{(out)}}_{\mu\nu} + \Theta(-L) g^{\textrm{(in)}}_{\mu\nu}$ and $\phi =\Theta(L) \phi^{\textrm{(out)}} +\Theta(-L) \phi^{\textrm{(in)}}$,
where `$\textrm{(out)}$' and `$\textrm{(in)}$' in the superscripts denote the outer and inner regions, respectively.
Then, the first junction conditions for the metric tensor and the dilaton field
can be imposed as \cite{Mann:1992my,Hyun:2006xt}
    \begin{gather}
    \label{eq:1st_junction_metric}
    \big[ g_{\mu\nu} \big] =0,\\
    \label{eq:1st_junction_dilaton}
    \big[ \phi \big]=0,
    \end{gather}
where $\big[ A \big]=A\big|_{L\to0+}-A\big|_{L\to0-}$.
Thus Eqs.~\eqref{eq:eom_total1} and \eqref{eq:eom_total2} can be rewritten as
    \begin{gather}
    \notag
    \Theta(L) e^{-2\phi^{\textrm{(out)}}} \left( R^{\textrm{(out)}}_{\mu\nu} + 2 \nabla^{\textrm{(out)}}_\mu \nabla^{\textrm{(out)}}_\nu \phi^{\textrm{(out)}} \right)
    + \Theta(-L) e^{-2 \phi^{\textrm{(in)}}} \left( R^{\textrm{(in)}}_{\mu\nu} + 2 \nabla^{\textrm{(in)}}_\mu \nabla^{\textrm{(in)}}_\nu \phi^{\textrm{(in)}} \right)\\ \label{eq:eom1}
    + \delta(L) e^{-2\phi} \left\{\frac 1 2 \left(n_\mu \kappa_{\nu\alpha}n^\alpha + \kappa_{\mu\alpha} n^\alpha n_\nu- \kappa_{\mu\nu}-\kappa_\alpha^\alpha n_\mu n_\nu \right) +2 n_\mu \big[ \partial_\nu \phi\big]\right\}
    = T^M_{\mu\nu}, \\
    \notag
    \Theta(L) \!\left(\! g^{\textrm{(out)}\mu\nu} R^{\textrm{(out)}}_{\mu\nu} \!-\! 4\left(\nabla^{\textrm{(out)}} \!\phi^{\textrm{(out)}} \right)^2 \!+\! 4 \Box^{\textrm{(out)}} \!\phi^{\textrm{(out)}} \!+\!\frac{4}{\ell^2} \!\right)\! \\
    \notag
    + \Theta(-L) \!\left(\! g^{\textrm{(in)}\mu\nu} R^{\textrm{(in)}}_{\mu\nu} \!-\! 4\left(\nabla^{\textrm{(in)}} \!\phi^{\textrm{(in)}} \right)^2+ 4 \Box^{\textrm{(in)}} \!\phi^{\textrm{(in)}} \!+\!\frac{4}{\ell^2} \!\right)\!\\ \label{eq:eom2}
    + \delta(L) \left(n^\alpha \kappa_{\alpha \beta} n^\beta - \kappa^\alpha_\alpha
    +n^\alpha \big[\partial_\alpha \phi \big] \right)=0.
    \end{gather}
Note that $\kappa_{\mu\nu}$ is defined through the relation of $\big[\partial_\gamma g_{\mu\nu} \big] = n_\gamma \kappa_{\mu\nu}$ with the unit vector $n_\mu$ which is normal to the hypersurface of $L=0$.
In Eq.~\eqref{eq:eom1}, the stress tensor can also be expressed as
    \begin{equation}
    \label{eq:em_tensor_total}
    T^M_{\mu\nu} = \Theta(L) T^{M\textrm{(out)}}_{\mu\nu} + \Theta(-L) T^{M\textrm{(in)}}_{\mu\nu} +\delta(L) T^{M\textrm{(edge)}}_{\mu\nu},
    \end{equation}
and so the dust cloud is described by choosing the stress tensor as
    \begin{equation}
    \label{eq:em_tensor_each}
    T^{M\textrm{(out)}}_{\mu\nu} = T^{M\textrm{(edge)}}_{\mu\nu} = 0,\qquad
    T^{M\textrm{(in)}}_{\mu\nu} = \rho u_\mu u_\nu ,
    \end{equation}
where $\rho$ is a proper energy density and $u^\mu$ is the three-velocity of the dust cloud.

From Eqs.~\eqref{eq:eom1}, \eqref{eq:eom2} and \eqref{eq:em_tensor_total}, the equations of motion in the outer spacetime of $L>0$
are obtained as
    \begin{gather}
    \label{eq:outer_eom1}
    e^{-2\phi^{\textrm{(out)}}} \left(R^{\textrm{(out)}}_{\mu\nu} + 2 \nabla^{\textrm{(out)}}_\mu \nabla^{\textrm{(out)}}_\nu \phi^{\textrm{(out)}}\right)=0,\\
    \label{eq:outer_eom2}
    g^{\textrm{(out)}\mu\nu} R^{\textrm{(out)}}_{\mu\nu} \!-\! 4\left(\nabla^{\textrm{(out)}} \!\phi^{\textrm{(out)}} \right)^2 \!+\! 4 \Box^{\textrm{(out)}} \!\phi^{\textrm{(out)}} \!+\!\frac{4}{\ell^2}=0,
    \end{gather}
and then, a static solution in terms of outer coordinates $(T,R,\sigma )$ is given as \cite{Horowitz:1993jc}
    \begin{equation}
    \label{eq:outer_codnt}
    ds^2_{\textrm{(out)}}  =- \left(1-\frac{M}{R} \right) dT^2+ \frac{\ell^2}{4 R^2} \left(1-\frac{M}{R} \right)^{-1} dR^2 +d\sigma^2,\quad
    \phi^{\textrm{(out)}} (R)= -\frac 1 2 \ln \left(R\ell\right),
    \end{equation}
where $M$ is a mass parameter of the black string and $R$ is a spatial coordinate which is $R\geq 0$.

In the inner spacetime of $L<0$, the equations of motion are
   \begin{gather}
    \label{eq:eom_inner1}
    e^{-2\phi^{\textrm{(in)}}} \left( R^{\textrm{(in)}}_{\mu\nu} + 2 \nabla^{\textrm{(in)}}_\mu \nabla^{\textrm{(in)}}_\nu \phi^{\textrm{(in)}}\right)= \rho u_\mu u_\nu,\\
    \label{eq:eom_inner2}
    g^{\textrm{(in)}\mu\nu} R^{\textrm{(in)}}_{\mu\nu} \!-\! 4\left(\nabla^{\textrm{(in)}} \!\phi^{\textrm{(in)}} \right)^2 \!+\! 4 \Box^{\textrm{(in)}} \!\phi^{\textrm{(in)}} \!+\!\frac{4}{\ell^2}=0,
    \end{gather}
and the line element and the dilaton field are assumed to be
    \begin{equation}
    \label{eq:ds_in}
    ds^2_{\textrm{(in)}} = -dt^2 +a^2 (t,r) dr^2 +d\sigma^2,\qquad \phi^{\textrm{(in)}}=\phi^{\textrm{(in)}}(t,r),
    \end{equation}
where $(t,r,\sigma)$ are comoving coordinates and $a(t,r)$ is a scale factor.
One might consider a more general form for interior metric.
For example, $g^{\textrm{(in)}}_{\sigma\sigma}$  is an arbitrary function of $t$ and $r$.
In fact, it would be smoothly matched on the dust edge,
but
we will assume the simplest form of $g^{\textrm{(in)}}_{\sigma\sigma}=1$
in order to find exactly soluble analytic solutions.
Note that the interior metric \eqref{eq:ds_in} would have the same topological
structure to the exterior metric \eqref{eq:outer_codnt}, since they should be smoothly matched
on the dust edge.

On the edge of $L=0$, the equations of motion can be read off
from the terms containing $\delta(L)$ in Eqs.~\eqref{eq:eom1}-\eqref{eq:em_tensor_total}.
Along the line of Refs.~\cite{poisson_2004,Ipser:1983db},
we can write the equations neatly as
    \begin{equation}
    \label{eq:eom_edge}
    -e^{-2\phi} \big[ K_{\mu\nu} \big] = h_\mu^\gamma T^{M\textrm{(edge)}}_{\gamma\nu},\qquad
    -2 \big[K_\alpha^\alpha \big]+ 4 n^\alpha \big[\partial_\alpha \phi \big] =0,
    \end{equation}
where $K_{\mu\nu} = h_\mu^\gamma \nabla_\gamma n_\nu$ is the extrinsic curvature of the hypersurface
and $h_{\mu\nu} = g_{\mu\nu}-n_\mu n_\nu$ is the projection operator onto the hypersurface.
The line element and the dilaton field on the edge are written as
    \begin{equation}
    \label{eq:ds_edge}
    ds^2_{\textrm{(edge)}} = -dt^2 +d\sigma^2,\qquad \phi^{\textrm{(edge)}}=\phi^{\textrm{(out)}} (\mathcal R (t))=\phi^{\textrm{(in)}}(t,r_0).
    \end{equation}
Note that the location of the edge can be expressed by
the inner coordinates as $r=r_0$ where $r_0$ is a constant
as well as the outer coordinates as $R=\mathcal{R} (t)$.

In the absence of matter on the edge, {\it i.e.}, $T^{M\textrm{(edge)}}_{\mu\nu}=0$,
the equations of motion \eqref{eq:eom_edge} lead to the second junction conditions \cite{Mann:1992my,Hyun:2006xt}
    \begin{align}
    \label{eq:2nd_junction_metric}
    \big[K_{\mu\nu} \big] & =0,\\
    \label{eq:2nd_junction_dilaton}
    n^\alpha \big[\partial_\alpha \phi \big] & =0.
    \end{align}
The normal vector in Eq.~\eqref{eq:2nd_junction_dilaton} can be written as
    \begin{equation}
    \label{eq:normal}
    n^\alpha =\left(\frac{\ell \dot{\mathcal R}}{2\mathcal R F(\mathcal R)},\frac{2\mathcal R F(\mathcal R) \dot T}{\ell},0\right),
    %\frac{\ell}{2\mathcal R} \big(-\dot{\mathcal R},\dot T,0\big),
    \qquad
    n^\alpha = \left(0,\frac{1}{a(t,r_0)},0\right)
    \end{equation}
in terms of the outer coordinates and the inner coordinates,
where $F(\mathcal R)=1-M/\mathcal{R}$ is the metric function in Eq.~\eqref{eq:outer_codnt} and
the overdot denotes the derivative with respect to the comoving time $t$,
and $\dot T$ can be obtained as
    \begin{equation}
    \label{eq:time_rel_1st_junction}
    \dot T = \frac{1}{F(\mathcal R )  }\sqrt{F(\mathcal R)+\frac{\ell^2\dot{\mathcal R}^2}{4\mathcal{R}^2}}
    \end{equation}
from the first junction condition \eqref{eq:1st_junction_metric}
which states that $ds_{\textrm{(out)}}^2 \big|_{R = \mathcal R (t)} =ds_{\textrm{(in)}}^2 \big|_{r = r_0}$.
From the second junction condition~\eqref{eq:2nd_junction_metric}, the only non-trivial component of $\big[ K_{\mu\nu} \big]$
is found to be
\begin{equation}
    \label{eq:Ktt}
    \big[ K_{tt} \big]= -\frac{2 \mathcal R}{\ell \dot{\mathcal R}} \frac{d}{dt} \sqrt{F(\mathcal R) +\frac{\ell^2 \dot{\mathcal R}^2}{4 \mathcal{R}^2}} =0.
    \end{equation}
Here, $\mathcal R (t)$ is assumed to be monotonic with respect to the comoving time
in order to avoid bouncing scenarios which are not of our interest, and thus
$(1/\dot{\mathcal R}) \left(d/dt\right)$ in Eq.~\eqref{eq:Ktt} is well-defined in such a way that it
can be replaced by $d/d\mathcal R$.
Then Eq.~\eqref{eq:Ktt} becomes
    \begin{equation}
    \label{eq:Ktt_R}
    -\frac{2 \mathcal R}{\ell} \frac{d}{d \mathcal R} \sqrt{F(\mathcal R) +\frac{\ell^2 \dot{\mathcal R}^2}{4 \mathcal{R}^2}} =0.
    \end{equation}
Using an integration constant defined by
    \begin{equation}
    \label{eq:eta_def}
    \eta^2 = F(\mathcal R_0 )+\frac{\ell^2 \dot{\mathcal{R}}^2_0}{4\mathcal{R}^2_0} ,
    \end{equation}
where $\mathcal{R}_0 = \mathcal R (0)$ and $\dot{\mathcal{R}}_0 = \dot{\mathcal R} (0)$,
we obtain the exact solution as
    \begin{equation}
    \label{eq:radius}
    \mathcal R (t) =
        \dfrac{M}{1-\eta^2} \sin^2 \left(\dfrac{\sqrt{1-\eta^2}}{\ell} (t-t_c)\right),
    \end{equation}
where a collapse time $t_c$ is defined by $\mathcal R (t_c) =0$ which gives \cite{Ross:1992ba,Mann:1992my}
    \begin{equation}
    \label{eq:collapsing_time}
    t_c =\dfrac{\ell}{\sqrt{1-\eta^2}} \sin^{-1} \sqrt{\dfrac{(1-\eta^2) \mathcal{R}_0 }{M}}.
    \end{equation}
For $|\eta|<1$, we need to impose the condition as $t_c \leq (\pi/2) (\ell/\sqrt{1-\eta^2})$
so that we can get a monotonically decreasing function, $i.e.$, $\dot{\mathcal R} <0$ in $0\leq t<t_c$.
Note that Eqs.~\eqref{eq:radius} and \eqref{eq:collapsing_time} also take the well-defined limit for $|\eta|\to 1$
as $\mathcal R (t)= M \big((t-t_c)/\ell\big)^2$ and $t_c=\ell\sqrt{\mathcal{R}_0/M}$.

From the second junction condition~\eqref{eq:2nd_junction_dilaton} with Eqs.~\eqref{eq:normal} and \eqref{eq:time_rel_1st_junction}, we obtain
    \begin{equation}
    \label{eq:dphi}
    \sqrt{\frac{F(\mathcal R)}{\ell^2}+\frac{\dot{\mathcal R}^2}{4\mathcal{R}^2}}+ \frac{\phi^{\textrm{(in)}}~\!\!'(t,r_0)}{a(t,r_0)} = 0,
    \end{equation}
where the prime denotes the derivative with respect to $r$.
Note that $\phi^{\textrm{(in)}}~\!\!'(t,r_0) \neq 0$ means that
the dilaton field in the inner spacetime would be spatially inhomogeneous so that the corresponding dust in that region
could also be inhomogeneous
\footnote[2]{If one were to consider homogeneous density and pressures,
then  equations of motions would be inconsistent with Eq.~\eqref{eq:dphi}.
As long as we require Eq.~\eqref{eq:dphi},
the inhomogeneity of a dust cloud is inevitable.}; otherwise, {\it i.e.},
$\phi^{\textrm{(in)}}~\!\!'(t,r_0) = 0$ means that $F(\mathcal{R}) < 0$
which leads to a strong restriction to the equation of motion and admits a
solution, starting to collapse only from inside the event horizon.
This fact was discussed in Ref.~\cite{Hyun:2006xt}, but
the explicit solutions for the inhomogeneous dust distribution in the inner spacetime has never been studied.
In the next section, we will find the collapsing solutions compatible with
the junction conditions \eqref{eq:1st_junction_metric}, \eqref{eq:1st_junction_dilaton},
\eqref{eq:2nd_junction_metric}, and \eqref{eq:2nd_junction_dilaton}.

\section{Analytic solutions to the dust cloud}
\label{sec:inner}
In order to obtain interior solutions smoothly matching the vacuum exterior \eqref{eq:outer_codnt},
we start with
the equations of motion
in the inner spacetime where $0\leq r \leq r_0$. From Eqs.~\eqref{eq:eom1}-\eqref{eq:em_tensor_total} in the comoving coordinates \eqref{eq:ds_in}, they are given as
    \begin{gather}
    \label{eq:inner_eom_a}
    e^{-2\phi^{\textrm{(in)}}} \left(-\frac{\ddot{a}}{a}+2\ddot \phi^{\textrm{(in)}} \right)= \rho,\\
    \label{eq:inner_eom_b}
    \dot{\phi}^{\textrm{(in)}}~\!\!' -\frac{\dot a}{a} \phi^{\textrm{(in)}}~\!\!'=0,\\
    \label{eq:inner_eom_c}
    \frac{\ddot{a}}{a}+ \frac{2}{a^2} \left( \phi^{\textrm{(in)}}~\!\!'' -a \dot a \dot{\phi}^{\textrm{(in)}} -\frac{a'}{a} \phi^{\textrm{(in)}}~\!\!' \right) =0,\\
    \label{eq:inner_eom_2}
    \left(\dot{\phi}^{\textrm{(in)}}\right)^2-\left(\frac{\phi^{\textrm{(in)}}~\!\!'}{a} \right)^2 -\ddot{\phi}^{\textrm{(in)}} +\frac{1}{\ell^2} =0.
    \end{gather}
Eq.~\eqref{eq:inner_eom_b} can be compactly written as $\partial_t (\phi^{\textrm{(in)}}~\!\!'/a) =0$, which
can be easily solved as
    \begin{equation}
    \label{eq:c1}
    \frac{\phi^{\textrm{(in)}}~\!\!'(t,r)}{a(t,r)} = \frac{c_1 (r)}{\ell},
     \end{equation}
where $c_1 (r)$ is an integration function.
Plugging Eq.~\eqref{eq:c1} into Eq.~\eqref{eq:inner_eom_2}, we obtain the differential equation as
    \begin{equation}
    \label{eq:simper_eom_phi}
    \left(\dot{\phi}^{\textrm{(in)}} (t,r)\right)^2-\ddot{\phi}^{\textrm{(in)}} (t,r) = \frac{c_1^2 (r) -1}{\ell^2}
    \end{equation}
whose general solution is obtained as
    \begin{equation}
    \label{eq:general_sol_phi}
    \phi^{\textrm{(in)}}(t,r) = -\frac 1 2 \ln \left[ c_3(r) \frac{\sin^2 \left( \sqrt{1-c_1^2(r)} \left(\dfrac t \ell + c_2 (r) \right) \right)}{1-c_1^2 (r)}\right],
    \end{equation}
where $c_2 (r)$ and $c_3(r)$ are integration functions.
Note that the solution \eqref{eq:general_sol_phi} is valid for any real value of $c_1 (r)$.
In particular, it has a smooth limit even in the limit of $c_1 (r) \to \pm1$ and its explicit
form is $\phi^{\textrm{(in)}}(t,r)=(-1/2) \ln \left[c_3(r) \left((t/\ell)+c_2(r) \right)^2 \right]$.

Let us now find an exact solution for $a(t,r)$.
We need to classify a set of solutions $\left( a, \phi^{\textrm{(in)}} \right)$
depending on $c_1(r)$ in Eq.~(32), and thus
we can consider largely two kinds of solution sets:
a first kind solution set (i) for $c_1(r)\neq 0$ and
a second kind solution set of (ii) for $c_1(r)= 0$.
In Eq.~(32), the first kind solution set (i) corresponds to a spatially inhomogeneous dilaton field
because $c_1(r)\neq 0$ indicates that $\phi^{\textrm{(in)}}~\!\!'(t,r)\neq 0$
while the second one (ii) corresponds to a spatially homogeneous dilaton field,
$i.e.$, $\phi^{\textrm{(in)}}~\!\!'(t,r) = 0$.
For each case, solving Eq.~\eqref{eq:inner_eom_c} with Eq.~\eqref{eq:general_sol_phi}, we obtain the scale factor as

(i) $c_1(r)\neq 0$
    \begin{align}
    \label{eq:case1_a}
    a(t,r)= & -\frac{\ell}{c_1(r)} \left\{
    \frac{c_1 (r) c_1 '(r)}{1-c_1^2 (r)}
    +\frac{c_3 '(r)}{2 c_3 (r)} \right.\\\notag
    & \left.
    + \left(
    -\frac{c_1(r) c_1 ' (r)}{\sqrt{1-c_1^2 (r)}} \left(\frac t \ell + c_2 (r) \right) +\sqrt{1-c_1^2 (r)} c_2 '(r)
    \right) \cot\left(\sqrt{1-c_1^2 (r)}\left(\frac t \ell + c_2 (r) \right)\right)
    \right\},
    \end{align}

(ii) $c_1(r)= 0$
    \begin{equation}
    \label{eq:case3_a}
    a(t,r) = c_4(r) \cot\left( \frac t \ell +c_2(r) \right)+c_5 (r).
    \end{equation}
In the case of (ii), $c_2(r)$ and $c_3(r)$ in Eqs.~\eqref{eq:general_sol_phi} and \eqref{eq:case3_a} must be constants to satisfy Eq.~\eqref{eq:c1} and $c_4 (r)$ and $c_5(r)$ are additional integration functions.
Also, note that,
in the limit of $c_1(r)  \to \pm1$ in Eq.~\eqref{eq:case1_a}, there exists a smooth limit as
$a(t,r) = \mp \ell \left[c_2 '(r) \big((t/\ell) +c_2 (r)\big)^{-1} +\right.$ $\left.c_3 '(r) \big(2 c_3 (r)\big)^{-1} \right]$.
%In particular, in the case of (ii), $c_1$, $c_2$ and $c_3$ must be constants in contrast to $c_4(r)$ and $c_5 (r)$.
From Eq.~\eqref{eq:inner_eom_a}, the energy density $\rho(t,r)$ for each case turns out to be
    \begin{align}
    \label{eq:inner_mass}
    \rho(t,r)  =\left\{
        \begin{matrix}
        -\dfrac{c_3 '(r)}{\ell{c_1(r)} a(t,r)} & \textrm{for~(i)}, \\
        \dfrac{2 c_3 (r)c_5(r)}{\ell^2 a(t,r)}  & \textrm{for~(ii)},
        \end{matrix}\right.
    \end{align}
which shows that $\rho(t,r) a(t,r)$ is time-independent.
This result is consistent with the covariant conservation law
of the stress tensor of the cylindrical dust clouds defined in Eq.~\eqref{eq:em_tensor_each}.

If we enforce the spatially homogeneous limit to the dilaton field in Eq.~\eqref{eq:general_sol_phi}
by taking $c_1(r)$ being non-zero constant and at the same time by choosing $c_2(r)$ and $c_3(r)$ to be some arbitrary
constants, then the scale factor \eqref{eq:case1_a} trivially vanishes.
Thus, we can conclude that the first kind solution set $\left( a(t,r), \phi^{\textrm{(in)}}(t,r) \right)$ is appropriate to describe the inhomogeneous dust distribution without
the homogeneous limit. On the other hand, as expected,
the solution set of Eqs.~\eqref{eq:general_sol_phi} and \eqref{eq:case3_a}
can be made spatially homogeneous
by taking $c_1(r)=0$
with $c_i(r)$ ($i=2,3,4,5$) being some constants.
Hence, the second kind solution set $\left( a(t,r), \phi^{\textrm{(in)}}(t) \right)$ can have a well-defined homogeneous limit such as $\left( a(t,r), \phi^{\textrm{(in)}}(t) \right)  \to \left( a(t), \phi^{\textrm{(in)}}(t) \right)$, which means that it can describe the inhomogeneous dust cloud as well as the homogenous one.
The homogeneous solutions in the second kind solution set are actually identical with the homogeneous cosmological solutions
in two-dimensional dilaton gravity \cite{Mann:1992yq},
once we identify the relevant constants as $c_2=\beta+(\pi/2)$, $c_3=e^{-2 \Phi_0}$, $c_4 =\lambda$ and $c_5 =\alpha$
\footnote[3]{The constants such as $\alpha$, $\beta$, $\lambda$ and $\Phi_0$ are defined in Ref.~\cite{Mann:1992yq}.}.
%The coordinate of $\sigma$ is decoupled from the three-dimensional cylindrically symmetric spacetime.

\section{Boundary conditions and physical requirements}
\label{sec:matching}
In this section, we fix the local unknown functions and their derivatives at $r=r_0$
by using the junction conditions along with some physical requirements.
Putting Eq.~\eqref{eq:radius} into Eq.~\eqref{eq:outer_codnt}, we find the dilaton solution on the edge
in terms of the outer coordinates as
    \begin{equation}
    \label{eq:phi_outer_explicit}
    \phi^{\textrm{(out)}} (\mathcal R (t))=-\frac 1 2 \ln \left[\frac{M\ell}{1-\eta^2} \sin^2 \!\bigg(\!\frac{\sqrt{1-\eta^2}}{\ell} (t-t_c)\!\bigg)\!\right].
    \end{equation}
From the first junction condition \eqref{eq:1st_junction_dilaton} for the dilaton field,
the dilaton field on the edge in terms of the inner coordinates is easily found to be
    \begin{equation}
    \label{eq:phi_inner}
    \phi^{\textrm{(in)}} (t,r_0) =-\frac 1 2 \ln \left[\frac{M\ell}{1-\eta^2} \sin^2 \!\bigg(\!\frac{\sqrt{1-\eta^2}}{\ell} (t-t_c)\!\bigg)\!\right].
    \end{equation}
Then, let us rewrite the second junction condition \eqref{eq:dphi} for the dilaton field
by using Eq.~\eqref{eq:eta_def} as
    \begin{equation}
    \notag
    \frac{\phi^{\textrm{(in)}}\!\!~' (t,r_0)}{a(t,r_0)} = -\frac{|\eta|}{\ell},
    \end{equation}
and thus $c_1(r)$ in Eq.~\eqref{eq:c1} can be determined as
    \begin{equation}
    \label{eq:bc1}
    c_1(r_0) =- |\eta|.
    \end{equation}
Comparing Eq.~\eqref{eq:general_sol_phi} at $r=r_0$
with the dilaton configuration \eqref{eq:phi_inner},
we get the following conditions:
    \begin{equation}
    \label{eq:bc2}
    c_2(r_0)=-\frac{t_c}{\ell},\qquad c_3(r_0) =M\ell.
    \end{equation}

Next, let us make an additional physical requirement in order to consider a smoothly collapsing dust.
The curvature scalar $R_{\rm curvature}$ on the edge in terms of the outer coordinates
\eqref{eq:outer_codnt} is calculated as
    \begin{equation}
    \label{eq:curv_out}
    R_{\rm curvature}(T,\mathcal{R}(t)) =
         \frac{4 (1-\eta^2)}{\ell^2 \sin^2 \big(\frac{\sqrt{1-\eta^2}}{\ell} (t-t_c)\big)},
    \end{equation}
where we used Eq.~\eqref{eq:radius}.
On the other hand, the curvature scalar can also be calculated by using
$R_{\textrm{curvature}}(t,r) = 2 \ddot a (t,r)/a(t,r)$ in terms of the inner coordinates \eqref{eq:ds_in}. Then, employing the boundary conditions  \eqref{eq:bc1} and \eqref{eq:bc2}, we can calculate the curvature scalar on the edge as
    \begin{equation}
    \label{eq:curv_in}
    R_{\textrm{curvature}}(t,r_0) =  \left\{
        \begin{matrix}
        \frac{4 (1-\eta^2)}{\ell^2 \sin^2 \left(\frac{\sqrt{1-\eta^2}}{\ell} (t-t_c)\right)} ~\dfrac{X}{X+\frac{(1-\eta^2) c_3'(r_0)}{2 M\ell c_1'(r_0)}} & \textrm{for~(i)}, \\
        \dfrac{4}{\ell^2 \sin^2 \left(\frac{t-t_c}{\ell}\right)} ~\dfrac{\tilde X}{\tilde X +c_5(r_0)} & \textrm{for~(ii)},
        \end{matrix}\right.
    \end{equation}
where
    \begin{equation*}
    X\!=\!-\!1\!+\!\sqrt{1\!-\!\eta^2} \!\left(\!\frac{t\!-\! t_c}{\ell} \!+\! \frac{(1 \!- \!\eta^2) c_2'(r_0)}{|\eta| c_1'(r_0)} \!\right)\! \cot\!\bigg(\!\!\frac{\sqrt{\!1-\!\eta^2}}{\ell} (t\!-\!t_c)\!\!\bigg)\!,~ \tilde X=c_4(r_0)\cot\left(\frac{t-t_c}{\ell}\right).
    \end{equation*}
Requiring the curvature scalars \eqref{eq:curv_out} and \eqref{eq:curv_in} to be the same on the edge
physically,
we can obtain
    \begin{align}
    \label{eq:bc3-1}
        \begin{matrix}
        c_3 '(r_0)=0 & \qquad\textrm{for~(i)}, \\
        c_5(r_0)=0 & \qquad\textrm{for~(ii)}.
        \end{matrix}
    \end{align}
In addition, in order to avoid the big rip of the infinite scale factor at $t=t_c$,
we also require
    \begin{align}
    \label{eq:bc3-2}
        \begin{matrix}
        c_2 '(r_0) =0 & \qquad\textrm{for~(i)}, \\
        c_4(r_0)=0 & \qquad\textrm{for~(ii)}.
        \end{matrix}
    \end{align}
Then, these conditions \eqref{eq:bc3-1} and \eqref{eq:bc3-2} result in
$a(t_c,r_0) =0$
which corresponds to $a(t_c) =0$ commonly defined in homogeneous cosmology \cite{Mann:1992yq}.

We will take the first kind solution set to describe the collapsing dust cloud
since $a(t,r_0)=0$ regardless of $t$ in the case of (ii).
Thus, the scale factor \eqref{eq:case1_a} at $r=r_0$
with the boundary conditions~\eqref{eq:bc1}, \eqref{eq:bc2}, \eqref{eq:bc3-1} and \eqref{eq:bc3-2}
is given as
    \begin{equation}
    \label{eq:add_cond_r0}
    a(t,r_0)=\frac{\ell c_1 '(r_0)}{1-\eta^2} \left\{\frac{\sqrt{1-\eta^2}}{\ell} (t-t_c) \cot \!\bigg(\!\frac{\sqrt{1-\eta^2}}{\ell} (t-t_c)\!\bigg)\! -1\right\}.
    \end{equation}
Finally, since the scale factor is positive during collapsing as
$a(t,r_0) >0$ in $0\leq t < t_c$, and it
leads to the following additional condition
    \begin{equation}
    \label{eq:bc4}
    c_1 '(r_0) <0.
    \end{equation}
This condition is responsible for the monotonically decreasing scale factor as $\dot a (t,r_0)<0$.

\section{Collapsing models for the inhomogeneous dust cloud}
\label{sec:simple}

%%%%%%%%%%%%%%%%%%%%%%%%%%%%%%
\begin{figure}[b]
\centering
\subfigure[]{\includegraphics[width=0.49\textwidth]{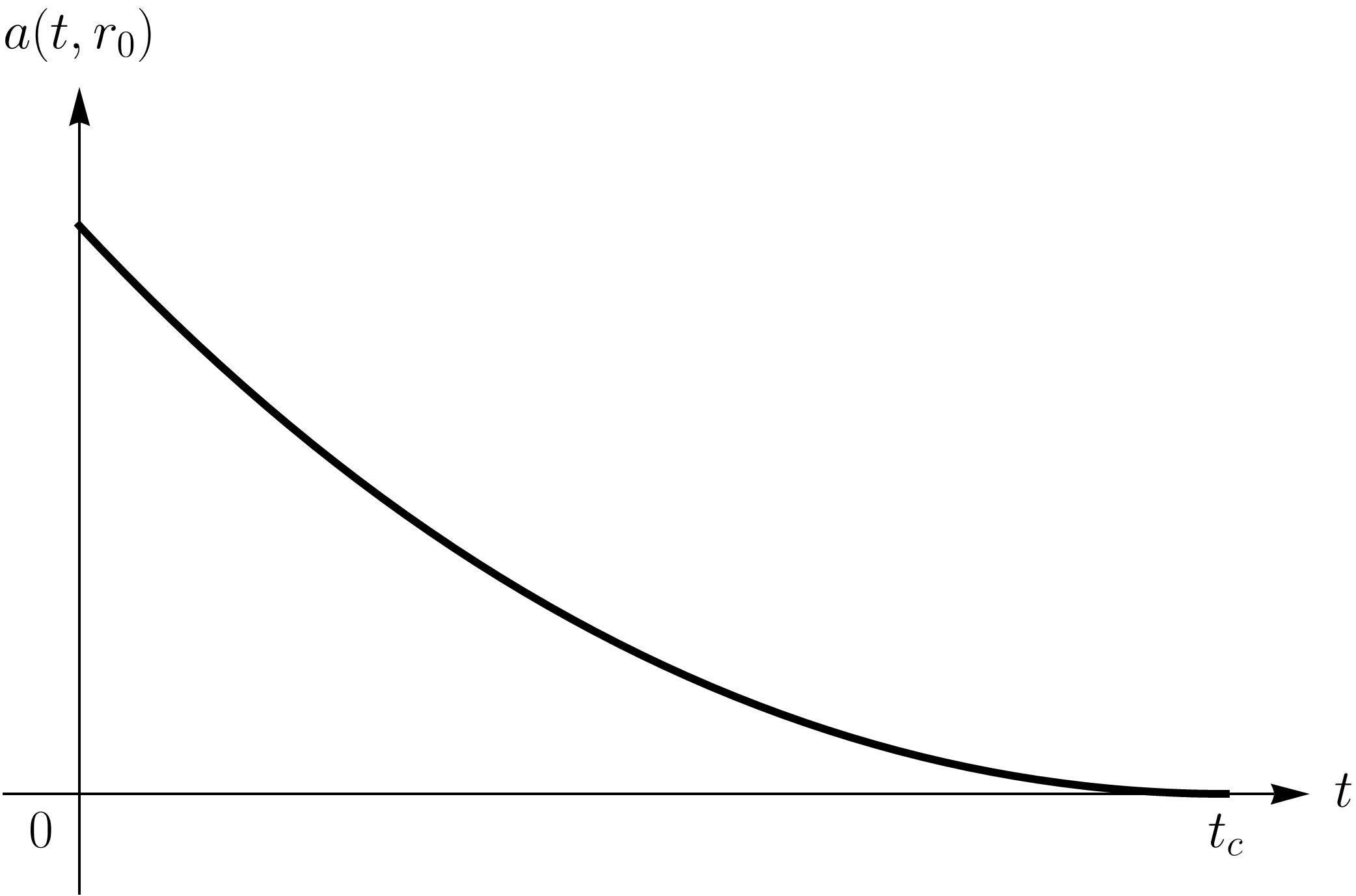}\label{fig:a}}
\subfigure[]{\includegraphics[width=0.49\textwidth]{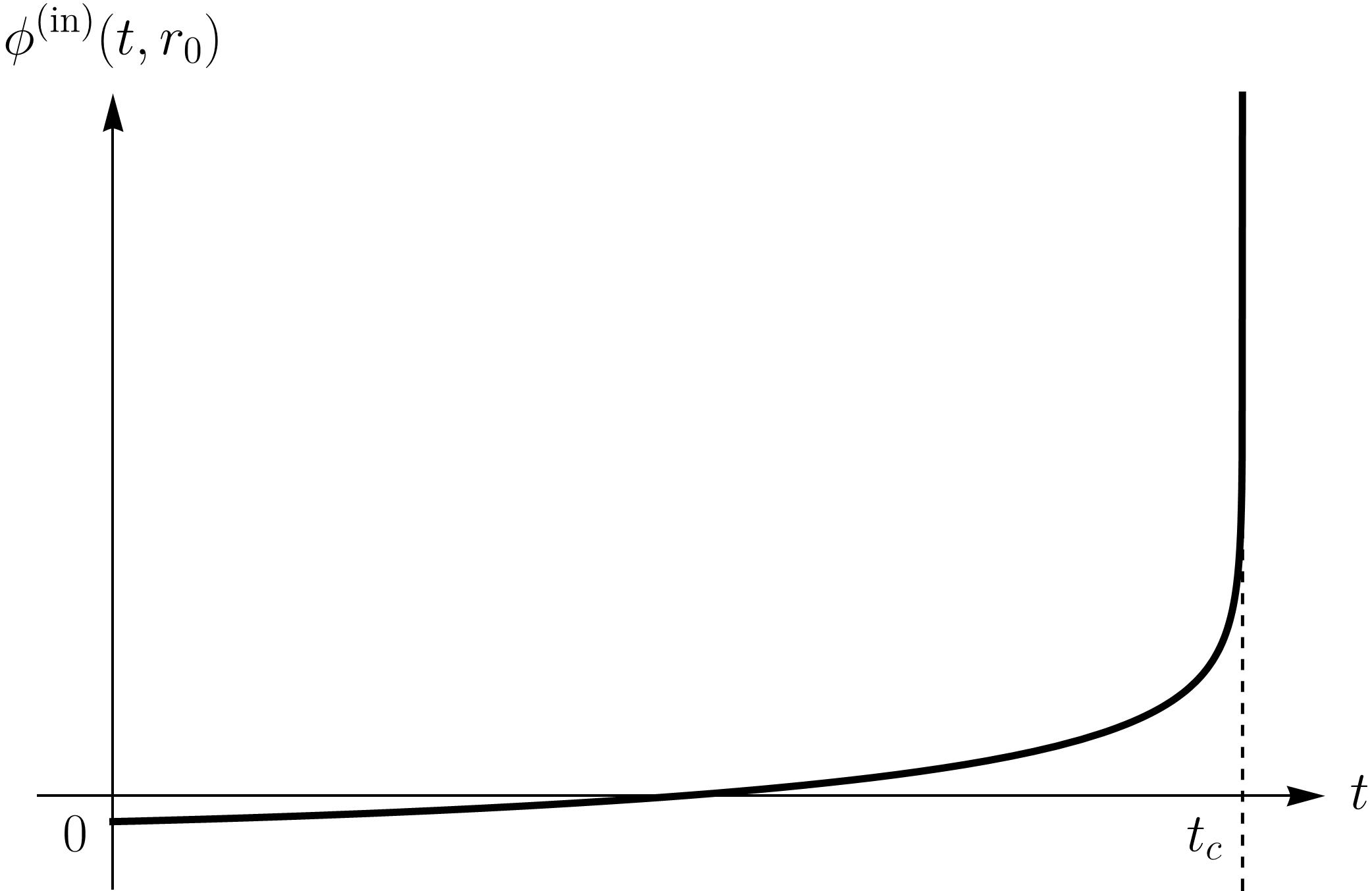}\label{fig:b}}
\caption{The decreasing scale factor \eqref{eq:final_sol_a} and the
increasing dilaton field \eqref{eq:final_sol_phi}
 on the edge with respect to the comoving time $t$ are plotted
by setting $\ell=r_0=1$, $M=2$, $\mathcal{R}_0 =3$ and $\eta=0.7$ for simplicity.
}%\label{fig:collapse}
\end{figure}
%%%%%%%%%%%%%%%%%%%%%%%%%%%%%%
We are now in a position to realize a collapsing model for the dust cloud
and so present one of the simplest dust models by choosing
    \begin{equation}
    \label{eq:simplest}
    c_1 (r) =-|\eta| \left(1 +\frac{\Delta}{2}\right),\quad
    c_2 (r) = -\frac{t_c}{\ell},\quad
    c_3 (r) = M\ell-M\ell |\eta| \Delta^2 \left(1+\frac \Delta 3 \right),
    \end{equation}
where $\Delta(r)=(r/r_0)-1$.
Note that the relations in Eq.~\eqref{eq:simplest} satisfy the junction conditions \eqref{eq:bc1} and \eqref{eq:bc2} as well as physical assumptions \eqref{eq:bc3-1}, \eqref{eq:bc3-2} and \eqref{eq:bc4}.
Consequently, the inhomogeneous solutions \eqref{eq:general_sol_phi} and \eqref{eq:case1_a} can be obtained explicitly in the form of
    \begin{align}
    \label{eq:final_sol_a}
    &a(t,r) = -\frac{\ell}{r_0} \frac{\Delta}{ 1-|\eta| \Delta^2 \left(1+\dfrac{\Delta}{3}\right)}\\
    \notag
    &+\frac{\ell |\eta|}{2 r_0} \left\{
    \frac{1-\sqrt{1-\eta^2 \left(1+\dfrac{\Delta}{2}\right)^2}\left(\dfrac{t-t_c}{\ell}\right)\cot
    \!\left(\!\sqrt{1-\eta^2 \left(1+\dfrac{\Delta}{2}\right)^2}\left(\dfrac{t-t_c}{\ell}\right)\!\right)\!
    }{1-\eta^2 \left(1+\dfrac{\Delta}{2}\right)^2}
    \right\},
    \end{align}
    \begin{align}
    \label{eq:final_sol_phi}
    &\phi^{\textrm{(in)}}(t,r) =-\frac 1 2 \ln\!\left[\!
    M\ell \!\left\{\!1-|\eta| \Delta^2 \left(1+\dfrac{\Delta}{3}\right)\!\right\}\!
    \frac{\sin^2
    \!\left(\!\sqrt{1-\eta^2 \left(1+\dfrac{\Delta}{2}\right)^2}\left(\dfrac{t-t_c}{\ell}\right)\!\right)\!
    }{1-\eta^2 \left(1+\dfrac{\Delta}{2}\right)^2}\right].
    \end{align}
At the collapse time, the scale factor vanishes on the edge while it is finite in the interior region,
and the dilaton field is divergent in the interior region as well as on the edge.
In particular, on the edge, the behaviors of the scale factor \eqref{eq:final_sol_a} and the dilaton field \eqref{eq:final_sol_phi}
with respect to the comoving time are plotted in Fig.~\ref{fig:a} and \ref{fig:b}, respectively.
The initial finite scale factor eventually vanishes
when the comoving time approaches the collapse time \eqref{eq:collapsing_time}
and the dilaton field diverges at the collapse time.
These results are consistent with the previous numerical analysis on the edge \cite{Hyun:2006xt}.

%On the other hand, the inhomogeneous dust cloud would have
%a singularity curve in $(t,r)$-plane \cite{Gutti:2005pn}.
From Eq.~\eqref{eq:final_sol_a}, the curvature scalar can be calculated as
    \begin{align} \label{deno}
    R_{\textrm{curvature}}(t,r)\!=\!
       \frac{2 |\eta|}{\ell r_0 a(t,r)} \frac{\!1-\!\sqrt{1\!-\!\eta^2 \!\left(\!1\!+\!\dfrac{\Delta}{2}\!\right)^2\!} \!\left(\!\dfrac{t\!-\!t_c}{\ell}\!\right) \cot\!\left(\!\sqrt{1\!-\!\eta^2 \!\left(\!1\!+\!\dfrac{\Delta}{2}\!\right)^2\!} \!\left(\!\dfrac{t-t_c}{\ell}\!\right)\!\right)\!}{\sin^2 \!\left(\!\sqrt{1\!-\!\eta^2 \!\left(1\!+\!\dfrac{\Delta}{2}\!\right)^2\!} \!\left(\!\dfrac{t\!-\!t_c}{\ell}\!\right)\!\right)\!},
    \end{align}
which is always finite for $0\leq r <r_0$ even at the collapse time $t_c$.
However, the curvature scalar \eqref{deno} on the dust edge reduces to
    \begin{equation}
    R_{\textrm{curvature}}(t, r_0)= \frac{4(1-\eta^2)}{\ell^{2} \sin^{2} \left(\sqrt{1-\eta^2} \left(\dfrac{t-t_c}{\ell}\right)\right)}
    \end{equation}
which is infinite as $t \to t_c$
\footnote[4]{A similar phenomenon also occurs for the interior BTZ solutions (disks of collapsing dust) \cite{Ross:1992ba,Gutti:2005pn}.
In the both of homogeneous and inhomogeneous dust clouds, the curvature scalars diverge when physical radii shrink to zero size,
which is originated from the fact that
the dust clouds collapse to a point (a string in our work) in finite proper time.}.
The curvature singularity occurs on the edge $(r=r_0)$, whereas it does not appear in the interior region ($0\leq r<r_0$).
This feature means that, as the comoving time approaches to the collapse time $t_c$,
although the curvature singularity cannot be detected by outside observers, inside the event horizon the spacetime is bipartized
into the two regular spacetimes around the edge.
Of course, the interior region is causally disconnected by this singularity,
so that any observer outside the edge fails to see this region.

%Note that there does not exist a time-like singularity
%even for $r_1= 2r_0 \left(|\eta|^{-1}-1\right)$ defined by $c_1(r_1)=-1$,
%since $R_{\textrm{curvature}}(t,r)\big|_{r\to r_1}$ is
%$4 \ell^{-2} \left[ \left((t-t_c)/\ell\right)^2 + 3c_3'/\left(2 c_1 c_1' c_3\right)\right]^{-1}$
%which diverges only at $t=t_c$.
%from Eqs.~\eqref{eq:case1_a} and \eqref{eq:simplest}.

Let us now calculate the total mass of the dust.
From the case (i) of our interest in Eq.~\eqref{eq:inner_mass} with the setting \eqref{eq:simplest},
we get the following relation of the energy density
    \begin{equation}
    \rho(t,r) a(t,r) =\frac{2 M}{r_0^2} \left(r_0 -r\right),
    \end{equation}
which results in
    \begin{equation}
    \label{eq:total_inner_mass}
    \int_0^{r_0} \rho(t,r) a(t,r)dr = M,
    \end{equation}
where the total mass of the dust cloud is related to the mass parameter of the black string.

Since our setting \eqref{eq:simplest} is actually not unique,
one may consider a slightly general model in terms of finite number of $\Delta$-polynomials by deforming only $c_3(r)$ such as
    \begin{equation}
    \label{newc3}
    c_3(r) = M\ell +|\eta| \ell \sum_{n=1}^{N} (-1)^n \alpha_n
    \left\{ \Delta^{n+1}  +\frac{(n+1)\Delta^{n+2}}{2(n+2)} \right\},
    \end{equation}
where $\alpha_n$'s are constants.
Of course, Eq.~\eqref{newc3} satisfies the previous boundary conditions of $c_3(r_0) =M\ell$ and
$c_3'(r_0)=0$.
The relation of $\rho(t,r) a(t,r)$ for the case (i) is calculated as
    \begin{equation}
    \rho(t,r) a(t,r)=\sum_{n=1}^N  \frac{(-1)^n (n+1) \alpha_n}{r_0} \Delta^n.
    \end{equation}
The total mass of the dust cloud should be preserved, and thus requiring Eq.~\eqref{eq:total_inner_mass},
the relation of $\alpha_n$'s can be obtained as
    \begin{equation}
    \label{eq:alpha_condition}
    \sum_{n=1}^{N} \alpha_n =M.
    \end{equation}
The combinations for $\alpha_n$'s are originated
from the wide variety of different configurations of the dust cloud.
In fact, within the same boundary conditions, there are infinitely many different solutions
depending on the inhomogeneous distribution of matter.

\section{conclusion and discussion}
\label{sec:conclusion}
In conclusion,
motivated by the formation of the BTZ black hole which is dual to the black string in three dimensions,
we studied the formation of the black string from the dust cloud.
The equations of motion for the inner spacetime describing the dust cloud were exactly solved for
arbitrary inhomogeneous distribution of the dust.
In order to smoothly match the inner spacetime and the outer spacetime of the black string,
the Israel junction conditions and
the physical requirements were imposed on the dust edge,
which tells us that
the inhomogeneous dust distribution is unavoidable.
Consequently, the scale factor on the edge was
shown to be monotonically decreasing with respect to the comoving time
and finally vanishes in the finite collapse time.
In addition, the curvature singularity occurs at the end of collapsing on the edge,
but they are cloaked inside the horizon of the black string.

One might wonder how to evade curvature and dilaton singularities
such as taking into account higher-curvature corrections.
In our work,
the curvature singularity and dilaton singularity occur on the edge.
In the low-energy effective action with higher-curvature corrections \cite{Edelstein:2018ewc},
the dilaton field of uncharged black string solution remains finite
although
the curvature scalar
still diverges when a physical radius goes to zero.
In this regard, we expect that the higher-curvature corrections could resolve
singularity problems in the formation of black string.

Let us discuss the T-duality of our interior solution.
The BTZ black hole and black string share the same equations of motion,
where they are related to the T-duality.
However, our interior solution is obviously not T-dual to
the interior BTZ solutions in Refs.~\cite{Ross:1992ba,Gutti:2005pn},
because equations of motion do not coincide.
Essentially,
in the presence of
the source term,
the T-dual transformation is no longer valid.
It would be interesting to extend the T-dual transformation incorporating the source terms,
which deserves further study.

As a comment,
the solutions \eqref{eq:general_sol_phi} and \eqref{eq:case1_a} in our model do not
have a continuous homogeneous limit
in contrast to the spherically symmetrical cases \cite{Ross:1992ba,Gutti:2005pn}.
One might wonder how to get the formation of the black string from the homogenous dust
which is also one of the realistic distributions of the dust.
We may find a clue from the previous work for the two-dimensional
dilaton gravity with a homogeneous dust \cite{Mann:1991ny},
where the authors took into account appropriate matter on the edge.
In our case also,
one may match the outer spacetime of the black string solution
and the inner spacetime of the homogeneous dust solution
by introducing additional junction matter, $i.e.$, $T^{M\textrm{(edge)}}_{\mu\nu} \neq 0$.
However, this homogeneous dust model would be quiet different from
the exactly soluble present model
since the existence of the junction matter
leads to the non-trivial modification of the second junction conditions \eqref{eq:2nd_junction_metric} and \eqref{eq:2nd_junction_dilaton}.
So this model may not ensure
analytically soluble equations of motion.

\acknowledgments
This work was supported by the National Research Foundation of Korea(NRF) grant funded by the Korea government(MSIT) (No. NRF-2022R1A2C1002894).
WK was partially supported by Basic Science Research Program through the National Research Foundation of Korea(NRF) funded by the Ministry of Education through the Center for Quantum Spacetime (CQUeST) of Sogang University (NRF-2020R1A6A1A03047877).
HE was partially supported by Basic Science Research Program (NRF-2022R1I1A1A01068833).

%%%%%%%%%%%%%%%%%%%%%%%%%%%%%%%%%%%%%%%%%%%%%%%%
%%%%%%%%%%%%%%%             References         %%%%%%%%%%%%%%%%

%\bibliographystyle{mybib}
%\bibliographystyle{apsrev4-1} % PRD
\bibliographystyle{JHEP}       %% JHEP.bst

\bibliography{references}

\end{document}